\documentclass[a4paper]{VisionStyle}
\usepackage{epsfig}

\begin{document}

\title{The Galactic Plane observed by XMM-Newton}

\author{A.\,Hands\inst{1} \and R.\,Warwick\inst{1} \and 
  M.\,Watson\inst{1} \and D.\,Helfand\inst{2} } 

\institute{
  Department of Physics and Astronomy, University of Leicester, LE1 7RH, UK
\and 
  Columbia Astrophysics Laboratory, New York, USA }

\maketitle 

\begin{abstract}

In AO-1 we proposed an ambitious long-term survey of selected regions
of our Galaxy (the XGPS survey) using the EPIC CCD
cameras on {\em XMM-Newton}. The first phase of this programme, which
aims to survey a strip of the Galactic plane in the Scutum region,
is currently underway. Here we report on the preliminary results
from the first $15$ survey pointings. We show that
the XGPS survey strategy of fairly shallow (5--10 ks) exposures
but wide-angle coverage is well tuned to the goal of providing a large 
catalogue of predominantly Galactic sources at relatively faint X-ray 
fluxes in the hard 2--6 keV band. 
 
\keywords{XMM-Newton: Galactic Plane: Serendipitous Surveys}
\end{abstract}

\section{Introduction}
  
With {\em XMM-Newton} we are able, for the  first time, to produce
high sensitivity, coherent surveys  of selected regions of our Galaxy.
The preliminary goal of  the {\em XMM-Newton} Galactic  Plane Survey 
(hereafter the XGPS survey) is to map a 5\degr~strip of the  Galactic Plane 
near Galactic longitude 22\degr. In addition to five XMM GT fields, a total of 
40 short observations amounting to 200 ks exposure time were 
awarded for this purpose in AO-1.  Here we report on the preliminary results
from this programme relating to nature of the Galactic X-ray source 
population and 
the origin of the bright X-ray emission which forms the so-called 
Galactic X-ray Ridge.

\section{The XGPS Source Catalogue}
\label{ahands-E1_sec:sdet}

\begin{figure*}[t]
\begin{center}
\epsfig{file=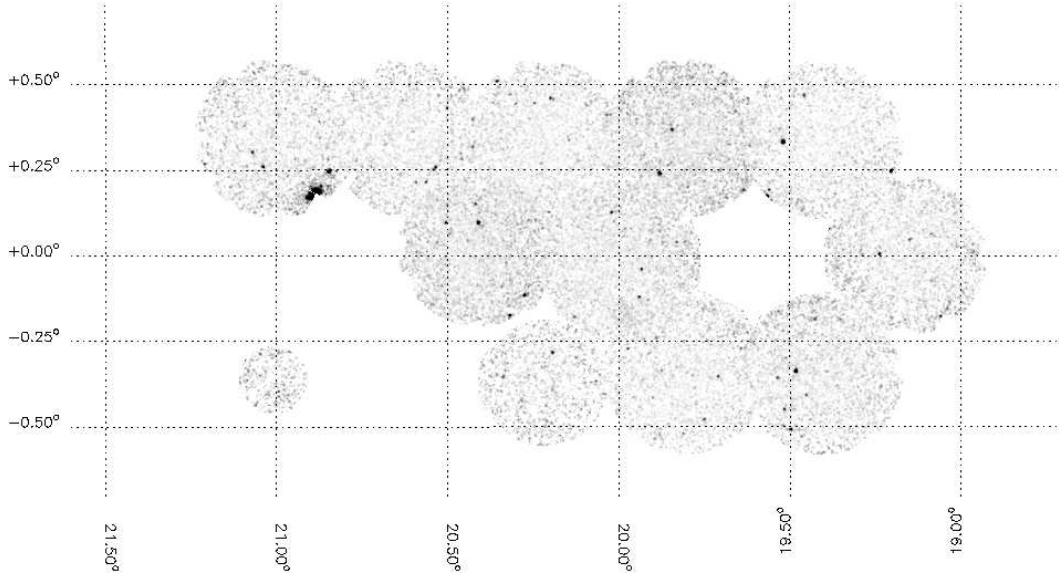, width=14cm}
\end{center}
\caption{ A mosaic of the available EPIC-MOS observations in the 2--6 keV 
band based on fields with a ``clean'' exposure time of more than 1 ks. 
Only a sub-set of sources detected in these observations are bright enough 
to be clearly visible in this image. }  
\label{ahands-E1_fig:fig1}
\end{figure*}

The  primary  purpose  of  the  XGPS survey  is to  study  the  X-ray  source
population of  the Galaxy at  faint fluxes.  To date $15$ pointings 
are available from the XGPS survey,  for which we have carried out 
source detection in  three  energy bands  -  soft (0.4--2  keV),  
hard  (2--6 keV)  and combined  (0.4--6 keV). In total 223  discrete 
sources have been detected at a significance level greater than 
5$\sigma$ in either the EPIC-pn and/or EPIC-MOS cameras.  
A subset of these sources can be seen in Fig. \ref{ahands-E1_fig:fig1} 
which  shows a mosaic of the hard-band images from the MOS 1/2  cameras.  
Only  one of these sources has a counterpart in the extensive, but low 
spatial resolution, survey of the Galactic Plane carried out by {\em ASCA} 
(\cite{ahands-E1:sug01}).  The brightest source (by approximately an
order  of  magnitude), which is clearly  visible  on  the eastern side
of the mosaiced region, does not appear in the {\em ASCA} catalogue and is
therefore a transient.

\subsection{The Transient Source}
\label{ahands-E1_sec:trans}

\begin{figure}[h]
\begin{center}
\epsfig{file=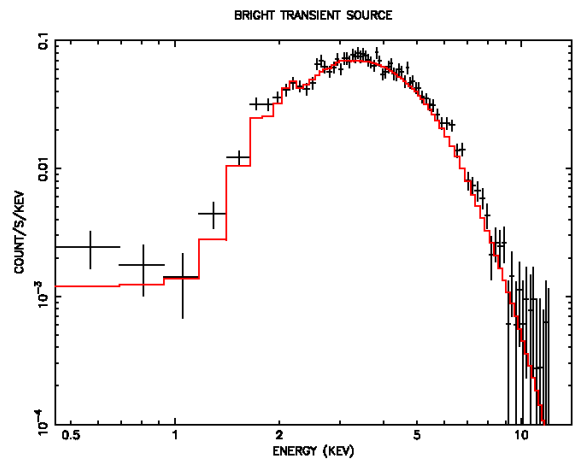, width=9cm}
\end{center}
\caption{The X-ray spectrum of the transient source in the XGPS\_009
field. }  
\label{ahands-E1_fig:fig2}
\end{figure}

The  bulk of  the  discrete X-ray sources  detected  in the XGPS survey  
are comprised of 
only a few tens  of counts.  However, in one of the survey fields
(XGPS\_009),  there is  a source, at  RA $\rm 18^h~28^m~34.0^s$,  
DEC -10\degr 36\arcmin 59\arcsec (J2000), which  contains  several thousand  
counts.   The X-ray spectrum  of this
object  is  shown  in  Fig.  \ref{ahands-E1_fig:fig2}.   Fitting  an
absorbed thermal  bremsstrahlung model yields  a temperature of $\sim 5$ keV
and an absorption column  of $\sim 6 \times 10^{22} \rm~cm^{-2}$.  This column 
density  is  consistent  with  either  an extragalactic  or  a distant  
Galactic origin.  In the latter case (assuming a distance 
of $\sim 15$ kpc) the observed flux is  equivalent to an X-ray luminosity 
of $\sim 10^{35} \rm~erg~s^{-1}$, suggesting the possibility that
this is a Be-star X-ray binary observed in a transient outburst state.

\subsection{Detection of SNR G20.0-0.2}
\label{ahands-E1_sec:g20}

\begin{figure}[h]
\begin{center}
\epsfig{file=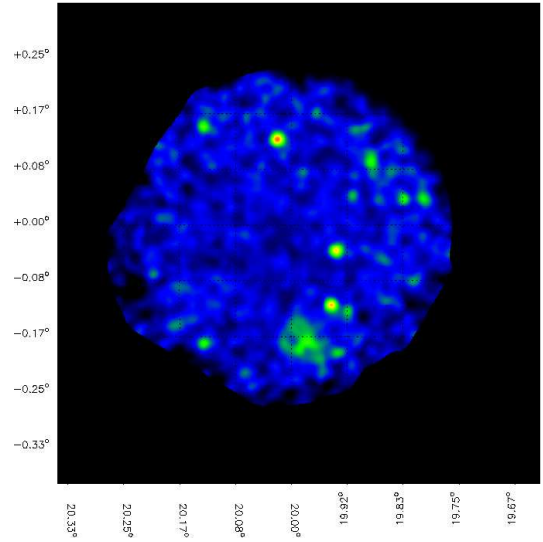, width=7cm}
\end{center}
\caption{The 2--6 keV pn plus MOS1/2 image of G20.0-0.2 in the Ridge 3 
field. }
\label{ahands-E1_fig:fig5}
\end{figure}

One important objective of the XGPS survey is to search for X-ray faint
supernova remnants (SNR)  with a view to placing improved constraints on  
the Galactic supernova rate and SNR evolution timescales. In this context 
we  have detected  an extended hard 
X-ray  source coincident with the known SNR  G20.0-0.2.  This  remnant,  
although well  documented through radio observations (\cite{ahands-E1:bh85}),
has not previously been observed in X-rays.  Fig. \ref{ahands-E1_fig:fig5}  
shows a smoothed image from one of the XGPS fields (Ridge\_3), 
based on the summation of the  pn and MOS 1/2 datasets.   The supernova  
remnant is  clearly visible  at the  bottom of the image; its 
X-ray flux is $\sim 3.5 \times 10^{-13} \rm~erg~s^{-1}~cm^{-2}$ (2--6 keV).

\begin{figure}[h]
\begin{center}
\epsfig{file=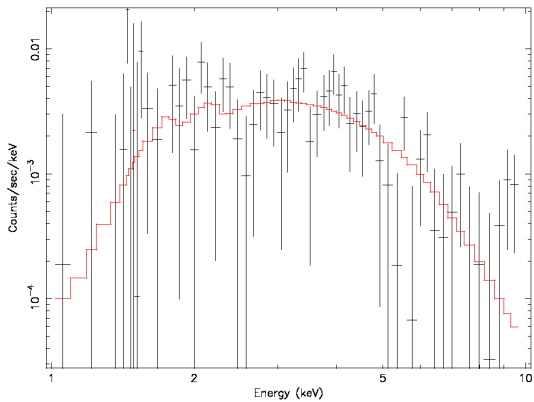, width=7cm}
\end{center}
\caption{The MOS spectrum of SNR G20.0-0.2.}
\label{ahands-E1_fig:fig6}
\end{figure}

The spectrum of the SNR may be found by using the whole of the Ridge\_3
field  (with  sources  removed)  as  a  background  template.   Fig.
\ref{ahands-E1_fig:fig6}  shows the resulting background subtracted
X-ray  spectrum.  We have fitted a simple absorbed power-law model to these
data; the derived parameter values are photon index $\Gamma = 2.1\pm0.7$ 
and absorption column density $\rm N_H = 4.4\pm1.8 \times 
10^{22}~cm^{-2}$.

The  fact  that  G20.0-0.2 has  not  been  seen  in X-rays  before  is
not surprising  given its relatively low X-ray surface brightness,
even though it appears to have a centre-filled X-ray morphology.   
Table \ref{ahands-E1_tab:tab1} compares the inferred radio and X-ray 
luminosities of G20.0-0.2 with those of the two Crab-like SNRs, 
3C58 and G21.5-0.9,  and the Crab itself. 
Although the radio  luminosity of G20.0-0.2 is on a  par with 3C58 and
G21.5-0.9, its X-ray luminosity is the  lowest of the four SNR by a 
factor of $\sim 2$ and is almost 4 orders of magnitude lower than that 
of the Crab nebula.

\begin{table}[h]
\caption{A comparison of the X-ray and radio luminosities of 4 SNRs. 
Here the X-ray band is defined as 0.1--4 keV and  radio band as 
10$^7$--10$^{11}$ Hz.}
\label{ahands-E1_tab:tab1}
\begin{center}
\leavevmode
\footnotesize
\begin{tabular}[h]{lccc}
\hline \\[-5pt]
SNR & L$_x$ & L$_R$ &  Log(L$_x$/L$_R$) \\
\hline \\[-5pt]
3C58 & $8.7 \times 10^{33}$ & $2.0 \times 10^{34}$ & -0.4 \\
G21.5-0.9 & $1.7 \times 10^{35}$ & $1.8 \times 10^{33}$ & 1.0 \\
Crab & $3.7 \times 10^{37}$ & $1.8 \times 10^{35}$ & 2.3 \\
G20.0-0.2 & $4.6 \times 10^{33}$ & $1.8 \times 10^{34}$ & -0.6 \\
\hline \\
\end{tabular}
\end{center}
\end{table}

\section{The Log N - Log S Function}
\label{ahands-E1_sec:log}

\begin{figure*}[t]
\begin{center}
\epsfig{file=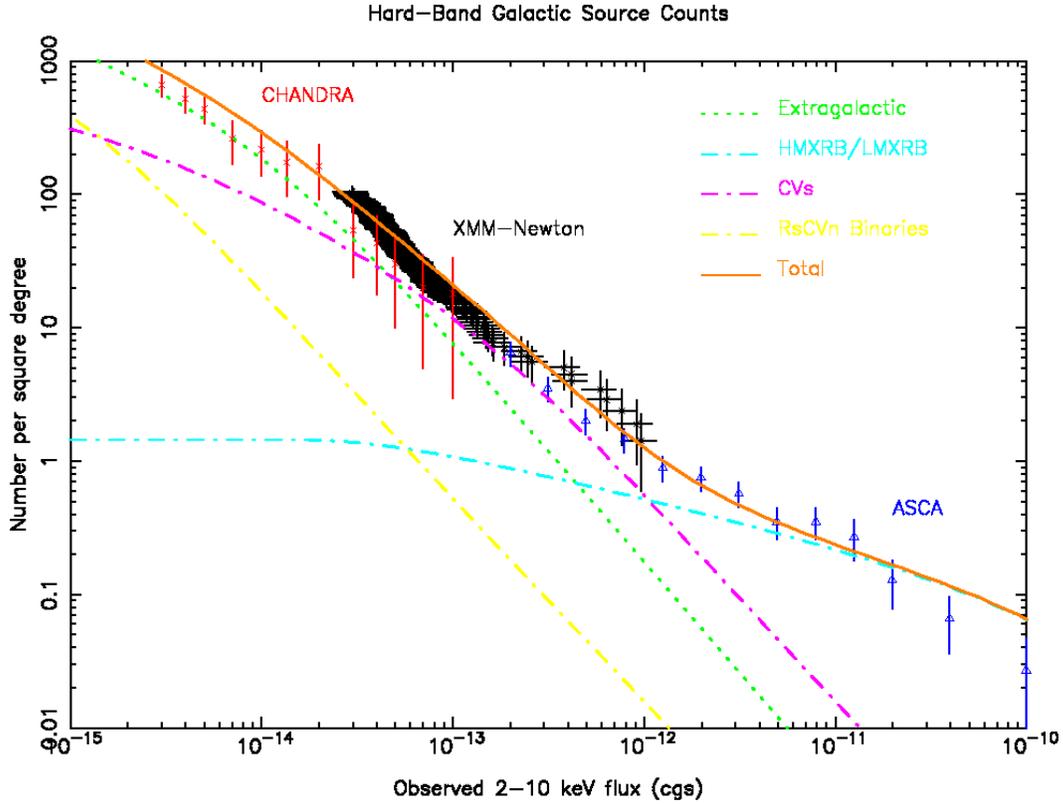, width=14cm}
 \end{center}
\caption{The hard (2--6 keV) band log N - log S plot based on 
XMM-Newton, Chandra and ASCA observations.  The curves 
show the predicted source counts for various Galactic source
populations.}  
\label{ahands-E1_fig:fig4}
\end{figure*}

We have used the X-ray source catalogue derived from the XGPS fields to
construct a log N - log S curve for the low Galactic latitude sky.
The normalisation and slope of this relation can, in principle,
provide important information  on the  spatial distribution and luminosity
function of the various Galactic source populations, albeit bound-up
with line-of-sight absorption effects.     Fig. \ref{ahands-E1_fig:fig4} 
shows  the log N - log S distribution, after correcting for sky coverage
effects, for {\em XMM-Newton} sources  detected in  the  hard (2--6 keV) 
band. For comparison equivalent results are also shown  for sources   
detected  by {\em ASCA} (\cite{ahands-E1:sug01})  and in recent deep
{\em Chandra}  (\cite{ahands-E1:ebi01}) observations. 
As can be seen in the figure, the flux range probed
by the XGPS  survey is intermediate between that achieved in the
{\em ASCA} and {\em Chandra}  programmes. The slope of  the 
log N - log S curve, over the full flux range sampled by the three
missions is, as expected, flatter than that measured for the extragalactic 
sky. Some comparisons with simple source distribution and luminosity
models are also shown (Fig. \ref{ahands-E1_fig:fig4}). Preliminary
analysis suggests that the Galactic population dominates these low
latitude counts down to a flux of $5 \times 10^{-14} \rm~erg~s~cm^{-2}$
(2--6 keV) below which the extragalactic contamination grows rapidly. 
Confirmation of this may be  found in the hardness ratio distribution
of the sources, which shows a wide spread of values, consistent
with the view that a highly absorbed  extragalactic population emerges
only at the lower end of flux range sampled in the XGPS survey.

This analysis demonstrates that the strategy of the XGPS survey, namely
relatively short (5--10 ks) observations but fairly wide-angle coverage
is rather well tuned to studying Galactic X-ray source populations at 
faint fluxes. 

\section{Optical Identification of XGPS Sources}
\label{ahands-E1_sec:opt}

A crucial step in the follow-up  of any X-ray survey is the identification
of the optical counterparts of the discrete sources detected in the survey. 
For one field in the XGPS region (the so-called Ridge\_3 field), a 
preliminary investigation of the optical counterparts has been
conducted through the auspices of the AXIS project (\cite{ahands-E1:bar02}).
Based on  sources derived from an earlier  source detection analysis,
\cite*{ahands-E1:mot02}  found that  out of  21 X-ray  sources  in the
Ridge\_3 field, 11 can be identified as stellar coronae from the
optical spectrum of a counterpart within the 90 \% error circle.  With
our current  source  detection  procedure  in the  three  energy  bands
(as detailed above) we now  find 29  source in  the Ridge\_3 field  at the
5$\sigma$ confidence  level (encompassing all but one  of the
original list  of 21 sources). 

We have investigated the correlation between  the hardness  of the 
X-ray  source and  the  brightness of the potential optical  counterpart.  
There is a clear  trend  for the  softer  X-ray  sources  to  have  
counterparts with  significantly brighter R-magnitudes  than those for
harder   sources.   Although  the   chance  of   coincidental  optical
correlations is not too high even at R=20, if we place a limit
at  $R \approx 17$, then this leaves  11 X-ray sources with potential 
(bright) counterparts. Of these, 8 have relatively soft X-ray spectra,
consistent  with the findings of  \cite*{ahands-E1:mot02} that these
are probably stellar coronal sources. Excluding a few
identified dMe stars, the bulk of the remaining $\sim 60\%$ of the sample,
including almost all the spectrally hard sources, are candidate
X-ray binaries and CVs, based on the source count analysis.
However, one of the sources is particularly interesting  because it is 
very  soft and yet has no visible optical counterpart to R$\approx$21. 
This is a candidate for an  isolated neutron star, although further 
analysis is needed to confirm this.

\section{Spectrum of the diffuse background}
\label{ahands-E1_sec:spec}

The  spectrum  of  the unresolved  Galactic  Ridge X-ray emission  (GRXE)  
was extracted  from the EPIC MOS observations by masking  out  the  known 
sources  and  summing  the
remaining data from all the observations.  The contribution of the
hard particle component of the EPIC background was estimated using
the edge regions of the MOS CCDs not illuminated by the sky ({\it i.e.,}
the regions of the detector which fall outside of the telescope field of 
view).  

\begin{figure}[h]
\begin{center}
\epsfig{file=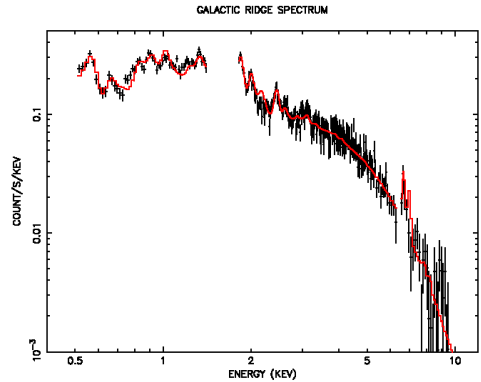, width=7.5cm}
\end{center}
\caption{The spectrum of the diffuse cosmic background in the Galactic 
Plane. This is the unresolved component of the Galactic X-ray Ridge.}
\label{ahands-E1_fig:fig3}
\end{figure}

Fig. \ref{ahands-E1_fig:fig3} shows   the  resulting  
background-subtracted   spectrum. 
A segment of the spectrum just below 2 keV 
is omitted due to the presence  of strong instrumental Al and Si 
fluorescent lines, which are  not uniformly distributed  across  
the  CCDs and hence are difficult to remove fully in the background
subtraction process.  We have fit this spectrum with   a
multi-temperature thermal (MEKAL) model (to account for emission from hot
Galactic plasma) and an absorbed  power-law continuum (representative
of  the extragalactic Cosmic X-ray Background). Best-fit  temperatures 
of $\sim0.1$, $\sim0.6$ and $\sim7$, keV were derived. Several emission 
lines visible  in the  softer part  of the  spectrum testify to the 
presence of the lower temperature components, whereas the iron K$_\alpha$
emission line detected  at 6.7 keV will be associated with the higher 
temperature plasma. The measured surface brightness in the 2--10 keV
band of the unresolved background was $\sim 10^{-10} 
\rm~erg~s^{-1}~cm^{-2}~deg^{-2}$. For comparison the integrated signal
in the source population resolved by {\em XMM-Newton} is  
$\sim 1.2 \times 10^{-11} \rm~erg~s^{-1}$ $\rm cm^{-2}~deg^{-2}$ 
for sources with fluxes between $3 \times 10^{-14}$ -- 
$10^{-12} \rm~erg$ $\rm s^{-1}~cm^{-2}$ (2--10 keV).

\section{Conclusions}
\label{ahands-E1_sec:disc}

The XGPS survey has  so far  yielded over  200 point
source detections  from 15 {\em XMM-Newton} observations. A realistic
target for the full AO1 programme is  therefore an X-ray source 
catalogue with between 500  and 1000 entries. This will 
provide a valuable resource for studying the Galactic
X-ray source population at faint fluxes ({\it i.e.,} down to
$F_X \sim 3 \times 10^{-14} \rm~erg~s^{-1}~cm^{-2}$).
The full XGPS survey  will also allow a detailed  analysis of the 
spectrum and distribution of the diffuse cosmic X-ray background 
on the Galactic plane. For example, the detection of spatial variations 
in the background intensity on scales of about a degree would provide
insight into origin and confinement mechanisms of the high temperature 
plasma which gives rise to the Galactic X-ray Ridge emission.

\end{document}